\documentclass[a4paper,twocolumn,11pt,accepted=2026-11-06]{quantumarticle}
\pdfoutput=1
\usepackage[utf8]{inputenc}
\usepackage[english]{babel}
\usepackage[T1]{fontenc}
\usepackage{amsmath}
\usepackage{hyperref}

\usepackage[numbers]{natbib}

\usepackage{tikz}
\usepackage{lipsum}
\definecolor{cset-aps-blueberry}{RGB}{28,128,158}
\definecolor{cset-aps-blue}{RGB}{46,44,184}
\definecolor{cset-aps-turquoise}{RGB}{0,67,88}
\definecolor{cset-aps-limegreen}{RGB}{190,219,67}
\definecolor{cset-aps-green}{RGB}{31,138,112}
\definecolor{cset-aps-yellow}{RGB}{255,225,25}
\definecolor{cset-aps-orange}{RGB}{253,116,0}
\definecolor{cset-aps-red}{RGB}{219,0,43}

\definecolor{blau}{HTML}{1575B9}
\definecolor{hellblau}{HTML}{65B7EF}
\definecolor{rot}{HTML}{E31B0A}
\definecolor{hellrot}{HTML}{FC6761}
\definecolor{gruen}{HTML}{25A131}
\definecolor{lila}{HTML}{2E2CB8}
\definecolor{grau}{HTML}{A6A6A6}

\hypersetup{
    colorlinks=true,
    linkcolor={cset-aps-red},
    linkbordercolor={cset-aps-red},
    filecolor={cset-aps-orange},
    filebordercolor={cset-aps-orange},
    citecolor={cset-aps-blue},
    citebordercolor={cset-aps-blue},
    urlcolor={cset-aps-green},
    urlbordercolor={cset-aps-green},
    menucolor={cset-aps-limegreen},
    menubordercolor={cset-aps-limegreen},
    breaklinks=true,
    pdfborderstyle={/S/U/W 2},
    pdfpagemode=UseOutlines,
    pdfstartpage={1},
}

\usepackage[utf8]{inputenc}
\usepackage[english]{babel}
\usepackage[T1]{fontenc}
\usepackage{amsmath}

\newcommand{\ii}{\text{i}}

\begin{document}

\title{Below-shot-noise capacity in phase estimation using nonlinear interferometers}

\newcommand{\affTUDaS}{
    \href{https://ror.org/05n911h24}{Technische Universit{\"a}t Darmstadt},
    Fachbereich Physik,
    Institut f{\"u}r Angewandte Physik,
    Schlo{\ss}gartenstr. 7,
    64289 Darmstadt,
    Germany
}

\newcommand{\affTUDaL}{
    \href{https://ror.org/05n911h24}{Technische Universit{\"a}t Darmstadt},
    Fachbereich Physik,
    Institut f{\"u}r Angewandte Physik,
    Otto-Berndt-Straße 3,
    64287 Darmstadt,
    Germany
}

\newcommand{\affICL}{
    \href{https://ror.org/041kmwe10}{Imperial College London},
    Department of Physics,
    1 Prince Consort Road,
    SW7 2AZ London,
    United Kingdom
}

\newcommand{\affUOX}{
    \href{https://ror.org/052gg0110}{University of Oxford},
    Department of Physics,
    Parks Road, 
    OX1 3PU Oxford,
    United Kingdom
}

\newcommand{\affIOF}{
    \href{https://ror.org/03g5ew477}{Fraunhofer Institute for Applied Optics and Precision Engineering IOF},
    Albert-Einstein-Str. 7, 
    07745 Jena, 
    Germany
}

\newcommand{\affICFO}{
    \href{https://ror.org/03g5ew477}{ICFO – Institut de Ciències Fotòniques},
    The Barcelona Institute of Science and Technology,
    08860 Castelldefels,
    Spain
}

\newcommand{\affUPoC}{
    Department of Signal Theory and Communications, 
    \href{https://ror.org/03mb6wj31}{Universitat Politecnica de Catalunya},
    08034 Barcelona,
    Spain
}

\author{Cristofero Oglialoro\,\includegraphics[width=7pt]{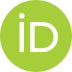}}
\affiliation{\affTUDaS}
\orcid{0009-0009-3162-1583}

\author{Gerard J. Machado\,\includegraphics[width=7pt]{orcid.png}}
\affiliation{\affICL}
\affiliation{\affUOX}
\orcid{0000-0002-0841-2195}

\author{Felix Farsch\,\includegraphics[width=7pt]{orcid.png}}
\orcid{0009-0001-2916-037X}
\affiliation{\affTUDaS}

\author{Daniel F. Urrego\,\includegraphics[width=7pt]{orcid.png}}
\affiliation{\affICFO}
\orcid{0000-0002-9067-7909}

\author{Alejandra A. Padilla\,\includegraphics[width=7pt]{orcid.png}}
\affiliation{\affICFO}
\orcid{0009-0007-7300-1716}

\author{Raj B. Patel\,\includegraphics[width=7pt]{orcid.png}}
\affiliation{\affICL}
\orcid{0000-0001-8627-1298}

\author{Ian A. Walmsley\,\includegraphics[width=7pt]{orcid.png}}
\affiliation{\affICL}
\affiliation{\affUOX}
\orcid{0000-0002-4714-0575}

\author{Markus Gräfe\,\includegraphics[width=7pt]{orcid.png}}
\affiliation{\affTUDaL}
\affiliation{\affIOF}
\orcid{0000-0001-8361-892X}

\author{Juan P. Torres\,\includegraphics[width=7pt]{orcid.png}}
\email{juanp.torres@icfo.eu}
\affiliation{\affICFO}
\affiliation{\affUPoC}
\orcid{0000-0002-4454-6676}

\author{Enno Giese\,\includegraphics[width=7pt]{orcid.png}}
\affiliation{\affTUDaS}
\orcid{0000-0002-1126-6352}
\email{enno.a.giese@gmail.com}

\begin{abstract}
Over the past decade, several schemes for imaging and sensing based on nonlinear interferometers have been proposed and demonstrated experimentally.
These interferometers exhibit two main advantages.
First, they enable probing a sample at a chosen wavelength while detecting light at a different wavelength with high efficiency (\emph{bicolor quantum imaging and sensing with undetected light}).
Second, they can show quantum-enhanced sensitivities below the shot-noise limit, potentially reaching Heisenberg-limited precision in parameter estimation.
Here, we compare three quantum-imaging configurations using only \emph{easily accessible} intensity-based measurements for phase estimation:
a Yurke‑type SU(1,1) interferometer, a Mandel‑type induced‑coherence interferometer, and a hybrid scheme that continuously interpolates between them. 
An ideal Yurke interferometer can exhibit Heisenberg scaling, however this advantage is known to be fragile under realistic detection constraints and in the presence of loss.
For a Mandel interferometer, we demonstrate that differential intensity detection in the unseeded, gain-balanced configuration provides the highest and most robust phase sensitivity, reaching, but not surpassing, the shot-noise limit with respect to the number of photons probing the sample, even in the presence of loss.
Intensity measurements in an unseeded Yurke-type configuration can achieve genuine sub-shot-noise sensitivity under balanced losses and moderate gain; however, their performance degrades in realistic high-gain regimes.
Consequently, in this regime, the Mandel configuration with differential detection outperforms the Yurke-type setup and constitutes the most robust approach for phase estimation.
\end{abstract}

\section{Introduction}
We consider here a class of optical interferometers that can be termed as nonlinear interferometers~\cite{chekhova2016} and which are based on original and fundamental ideas developed some decades ago~\cite{yurke1986,zou1991}.
This type of interferometers uses optical parametric amplifiers at the input and output ports in place of the beam splitters of a standard interferometer.
These interferometers have gained increasing attention~\cite{Hochraine2022,fuenzalida2024a} as a useful tool for imaging, parameter estimation, spectroscopy and optical coherence tomography, to name a few applications.
They offer both practical advantages~\cite{lemos2014,fuenzalida2023,valles2018} and a way to achieve quantum-enhanced sensitivity~\cite{giese2017,kranias2025,florez2018}, i.\,e., going below the shot-noise limit in parameter estimation.

\begin{figure}[t]
  \centering
  \includegraphics[width=\columnwidth]{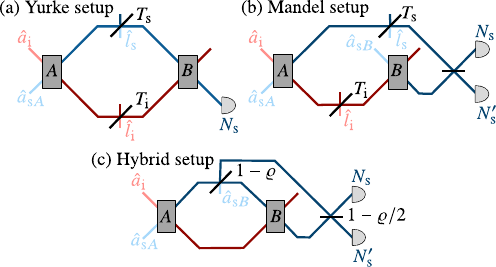}
  \caption{
    Sketches of different nonlinear interferometers:
    In the Yurke setup (a), a nonlinear medium $A$ with vacuum input generates signal and idler fields $j=\text{s},\text{i}$ (blue and red).
    They may be subject to loss, encoded in a non-unit transmittance $T_j$, before they seed a second nonlinear medium $B$, whose signal output $N_\text{s}$ is detected.
    In the Mandel setup (b), only the idler seeds medium $B$, while the signal modes of both media are interfered on a 50:50 beam splitter, whose outputs $N_\text{s}$ and $N_\text{s}^\prime$ are detected.
    In a hybrid configuration (c), only a fraction $1-\varrho$ of the signal mode is seeded into medium $B$, while the remaining part is interfered with the signal output of medium $B$ on a beam splitter, whose transmittance is $1-\varrho/2$.
    The variable parameter $\varrho$ tunes the setup from a Yurke ($\varrho=0$) to a Mandel configuration ($\varrho=1$).
  }
  \label{fig:setups}
\end{figure}

We consider three different types of nonlinear interferometers, namely the Yurke, Mandel, and hybrid schemes, shown in Fig.~\ref{fig:setups}.
We will  consider the case of unseeded configurations, where the input quantum state of imaging photons is the vacuum.
In all three cases, pumping the nonlinear medium $A$ with monochromatic light, alongside vacuum input, induces nondegenerate parametric down-conversion. 
The medium generates $n$ photons in each of two modes, called signal (blue) and idler (red) fields.
The number of generated photons depends on the parametric gain, which is determined mainly by the intensity of the pump field and the nonlinear susceptibility of the medium~\cite{torres2011}.

In the remainder of this article, we use this photon number as the resource for defining shot-noise scaling $1/\sqrt{n}$ as a benchmark and refer to $1/n$ as Heisenberg scaling. 
Thus, all scaling statements refer to the number of photons probing the sample, while noting that the gain of the first crystal provides only one possible figure of merit and that alternative resources may be considered~\cite{giese2017}.

Formally, the light generation process during parametric down-conversion can be described by a Bogoliubov transformation, whose coupling matrix is an element of the SU(1,1) group~\cite{biedenharn1965}.
Due to the nondegenerate nature of the process, the signal and idler fields may be at different wavelengths, enabling bicolor sensing with undetected photons.
For example, the signal mode might be in the visible range of the spectrum and the idler mode in the mid-infrared.
The object is probed by the idler light and has a complex transmission coefficient $t_\text{i}=\sqrt{T_\text{i}} \exp(\ii \varphi_\text{i})$, with transmittance $0\leq T_\text{i}\leq 1$ and a phase $\varphi_\text{i}$.
In the transmittance we may also include other loss channels of the idler arm.
Loss in the signal arm is included through a transmission coefficient  $t_\text{s}=\sqrt{T_\text{s}} \exp(\ii \varphi_\text{s})$, with transmittance $0\leq T_\text{s}\leq 1$, the phase shift experienced by the signal is $\varphi_\text{s}$.
In general, the signal/idler transmittances may be different.

All three interferometers begin with a two‑mode‑squeezed vacuum interaction generated in medium $A$, which produces correlated signal–idler beams, each with $n$ photons. 
In the ideal lossless case, the quantum Fisher information for estimating the phase $\phi=\varphi_s+\varphi_i$, based on the joint quantum state of signal/idler photons generated in the first nonlinear crystal, is $F_Q=4 n (n+1)$.
This is a measure of the optimal sensitivity achievable with any measurement.
This bound applies to both the Yurke and Mandel configurations and implies the possibility of reaching the Heisenberg limit of precision ~\cite{giese2017,kranias2025}.

Although the initial squeezing process and corresponding quantum Fisher information are common to all three configurations, the evolution of the global quantum state differs in each setup. 
In the Yurke~\cite{yurke1986} setup shown in Fig.~\ref{fig:setups}\,(a) both signal and idler modes are seeded into a second nonlinear medium $B$.
The ideal output quantum state contains thus two modes, signal and idler.
The output signal mode intensity ($N_\text{s}$) is detected.
Hence, two-mode squeezed state generated in medium $A$ is either further squeezed/amplified by medium $B$ or antisqueezed/deamplified~\cite{herzog1994}.
Because squeezing is a phase-sensitive process that depends on the phases of the pump, signal, and idler beams, these phases determine the detected intensity.

In the Mandel setup shown in Fig.~\ref{fig:setups}\,(b), only the idler mode is seeded into the second nonlinear crystal, so that the output quantum state comprises three modes (two signal modes and one idler mode).
When considering all modes in the Yurke or Mandel setups, the quantum Fisher information remains the same as the initial one stated above, since, in the lossless case, the corresponding transformations are unitary \cite{kranias2025}.
However, when restricting measurements to signal modes only, the quantum Fisher information of the reduced quantum state, obtained by tracing out the idler mode, differs for the Yurke and Mandel setups.

In addition to the gains in the two nonlinear media, the sensitivity of phase estimation achievable in a Yurke configuration depends on the losses in the signal and idler arms of the interferometer being equal (balanced) or not (unbalanced). This effect is a manifestation of the strong dependence of the degree of squeezing on the presence of loss.
Indeed, for a particular value of the gain in the low-gain regime, unbalanced loss can even provide a better sensitivity compared to a balanced situation~\cite{Michael2021}.
Complementing these results, we demonstrate below that balanced and unbalanced situations in a Yurke configuration lead to radically different behaviors of the sensitivity in the high-gain regime.

In principle, a loss-induced degradation of sensitivity can be partially compensated by using different gains~\cite{giese2017,manceau2017} in media $A$ and $B$.
The objective of this work is to compare detection schemes under identical interferometric conditions. 
Gain optimization in the presence of loss represents an additional degree of freedom and requires a careful reassessment of the relevant resource for sensitivity comparisons~\cite{giese2017}, in addition to posing experimental challenges~\cite{manceau2017a} under {\it realistic} scenarios. 
A systematic analysis of this optimization is beyond the scope of the present study and we restrict ourselves to the case of equal gains for media $A$ and $B$.

It is well-known that squeezed states can enable an interferometer to achieve a sensitivity below shot noise, which in the ideal case is such that the phase uncertainty has a Heisenberg $1/n$-scaling.
Moreover, seeding both idler and signal into medium $B$ gives rise to a compact and symmetric form, which is robust for many applications like spectroscopy~\cite{Lindner2021,Tashima24}, microscopy~\cite{Kviatkovsky2020,Gilaberte2021,leon-torres2026}, and holography~\cite{topfer2022,leon-torres2024,Topfer2025,krause2026}.
However, under loss, the squeezed quantum state deteriorates quickly~\cite{marino2012,Gong2017}, losing its sub-shot-noise capability.

In contrast, in the Mandel~\cite{wang1991} setup shown in Fig.~\ref{fig:setups}\,(b), only the idler mode seeds medium $B$. 
The signal modes emitted by media $A$ and $B$ interfere on a 50:50 beam splitter.
The intensities measured at the two exits of the beam splitter, $N_\text{s}$ and $N_\text{s}^\prime$, depend on the phases of the pump, signal, and idler beams inside the interferometer.
However, because the nonlinear process in medium $B$ is partially seeded, the intensities of the two signal modes can differ significantly~\cite{wiseman2000,kolobov2017, machado2024}.
In fact, such an interferometer does not display a pure SU(1,1) symmetry, because there are three input and output modes.

In the Yurke configuration, an intensity measurement of its signal output at the dark fringe already saturates this bound.
In contrast, in the Mandel setup, the quantum Fisher information of the signal modes scales like shot noise~\cite{kranias2025}, and no measurement on the signal modes alone, including homodyne detection, can restore Heisenberg scaling.
Instead, to extract a phase with such a high precision requires a joint measurement of signal and idler modes with much more involved experimental methods like homodyning, similar to truncated SU(1,1) interferometers~\cite{anderson2017,anderson2017a,gupta18,caves2020}.

For the Mandel configuration, we use a Gaussian uncertainty analysis of the intensity measurement as a metric for the phase sensitivity.
For high gain, which is the most interesting regime for quantum metrology, signal modes of different intensity interfere, such that the contrast of $N_\text{s}$ decreases significantly~\cite{wiseman2000,kolobov2017, machado2024} with adverse consequences for the sensitivity.
Here, we demonstrate below that the intensity difference of both exits~\cite{miller2021}, namely $N_-=N_\text{s}-N_\text{s}^\prime$,  provides a robust estimator against this deleterious effect and in fact scales like shot noise.

A comparison of the Yurke and Mandel configurations~\cite{Roeder2025}, based on the quantum Fisher information~\cite{kranias2025, Houde2026} and intensity as a phase estimator~\cite{miller2021}, suggests different implications for the attainable phase sensitivity. 
These previous studies have explored both configurations with additional coherent seeds in the signal and/or idler modes, and the comparison has been extended experimentally to the low-gain regime~\cite{moos2026}. 
In our comparative study, however, we focus on the vacuum-input scenario to establish a well-defined baseline for the comparison of the different configurations and facilitate a clear discussion of the underlying physics.

We also discuss a hybrid setup that has been already implemented in the low gain regime~\cite{gemmell2024}. This scheme is sketched in Fig.~\ref{fig:setups}\,(c).
A beam splitter with transmittance $1-\varrho$ couples a fraction of the signal beam out of the Yurke configuration.
This fraction interferes with the signal mode emitted by medium $B$ on a flexible beam splitter with transmittance $1-\varrho/2$.
Such a configuration allows for tuning from the Yurke case ($\varrho =0$) to the Mandel case ($\varrho =1$) and studying the transition between both configurations.

Here, we analyze the phase estimation sensitivity of the different configurations by assuming equal gain for both nonlinear media and by approximating the signal and idler modes as single-mode beams.
Despite its simplicity, this approach provides useful and qualitatively accurate results.
Heuristically, one can think of $N_\text{s}$ as the number of signal photons generated per unit bandwidth ($\textrm{photons}/\textrm{s}/\textrm{Hz}$), so that $N_\text{s}$ can be related to the flux rate $R_\text{s}= N_\text{s}  \times \mathcal{B}$ of signal photons observed in an experiment, where $\mathcal{B}$ is the spectral bandwidth of detection.
More rigorously, it can be demonstrated that the validity of this result assumes a single transverse mode and a detection bandwidth $\mathcal{B}$ much smaller than the bandwidth of parametric down-conversion~\cite{harris2005} in media $A$ and $B$.
For the variance, one can derive a similar expression, even when one includes the spatial/frequency multimode character of signal/idler modes~\cite{gene2006}.

\section{Phase uncertainty: the lossless case}
\label{sec:lossless}
We begin our discussion by studying these configurations with no input but the pump field, equal gain in both nonlinear media, and without any loss.
The mathematical expressions of the interference patterns and variances are derived in Appendix~\ref{app:Interference_patterns}.
We find that detecting only one exit of the (Mandel or Yurke) interferometer, the signal photon number $N_\text{s}$ depends on the phase $\phi$ to be determined and follows thermal statistics as expected for squeezed light.
Therefore, it has a variance $\text{Var} \left( N_\text{s} \right)= N_\text{s}  \left[ 1+ N_\text{s} \right]$.
These statistics imply through Gaussian uncertainty propagation that the phase uncertainty for the Yurke (Y) configuration and the uncertainty observed for detecting a single exit of the Mandel configuration (SM) takes the form
\begin{align}
\label{eq:sigma_SM/Y}
    \sigma^2_\text{Y/SM} = \frac{ N_\text{s}  \left[ 1+N_\text{s} \right]}{\left( \partial N_\text{s}/\partial \phi \right)^2}.
\end{align}
In the Yurke configuration, the interference pattern takes the form $N_\text{s}=2 n (1+n)(1+\cos\phi)$, where $n$ corresponds to photons generated in nonlinear medium $A$ and depends on the parametric gain.
Furthermore we find that in the case of balanced gain and no losses, when all light generated in $A$ is converted back to the pump in $B$, the interference pattern shows perfect contrast.

Because the statistics are thermal, the variance vanishes at destructive interference, namely for $\phi=\pi$.
Therefore, it is possible to work at the dark fringe with minimal intensity fluctuations.
Taking this limit, we find~\cite{yurke1986}
\begin{align}
\label{eq:perfect_Yurke_Scaling}
     \left.\sigma^2_\text{Y}\right|_\pi =  \frac{1}{4n(n+1)} \overset{\text{hg}}{\longrightarrow} \frac{1}{4n^2},
\end{align}
where we have expanded the result in the last step in orders of $1/n$ and observe genuine Heisenberg scaling in the high-gain (hg) regime $1 \ll n$.
One can show that this expression saturates the quantum Cram\'er-Rao bound and therefore constitutes an optimal measurement, which implies that the mean photon number obtained from an intensity measurement is the optimal estimator for the interferometer phase~\cite{kranias2025}.

In contrast, the interference pattern $N_\text{s}= n (n + 2 + \sqrt{n+1} \cos\phi) /2$ observed in one of the two exits of the Mandel configuration does not display perfect destructive interference for arbitrary $n$, so that the thermal variance never vanishes.
Hence, there is no benefit from working at the dark fringe and the working point of highest sensitivity is obtained for a different phase.
Details on the optimization can be found in Sec.~\ref{sec:loss_Yurke}.
After this procedure, we find that the minimal phase uncertainty
\begin{align}
\label{s1}
   \left. \sigma_\text{SM}^2\right|_{\phi_\text{min}} =\frac{1}{4n} \frac{n+2}{n+1}+\frac{n^2 + \sqrt{4 (n+1)^2 + n^4}}{8(n+1)},
\end{align}
not only scales worse than shot noise, but even deteriorates in the high-gain regime, as $ \sigma_\text{SM}^2|_{\phi_\text{min}}\!\!\overset{\text{hg}}{\longrightarrow} n/4$.
This behavior is a consequence of the fact that photon-number measurements do not necessarily constitute optimal measurements.

\begin{figure}[t!]
    \centering
    \includegraphics[width=1\linewidth]{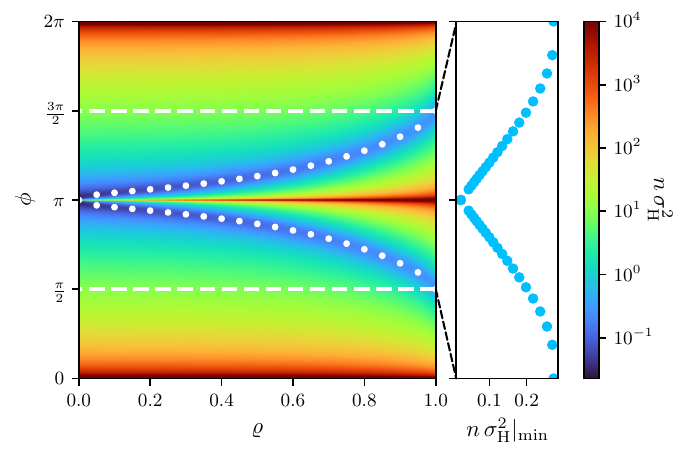}
    \caption{
    Phase uncertainty of the hybrid setup $n\,\sigma^2_\text{H}$ for $n=10$ and equal gain in both nonlinear media, as a function of the mixing parameter $\varrho$ and the phase $\phi$. 
    The phases minimizing the sensitivity for discrete values of $\varrho$ are represented by white dots. The optimal working point for the Yurke setup ($\varrho=0$) is at $\phi_\text{min}=\pi$, while the sensitivity is minimized at two phases for $\varrho>0$.
    For the Mandel setup ($\varrho=1$), these phases are $\phi_\text{min} = \pi/2,\,3\pi/2$.
    The margin shows the minimal phase uncertainties, ranging from the optimal uncertainty of the Yurke setup to the shot-noise limit exhibited by the Mandel configuration.
    }
    \label{fig:hybrid}
\end{figure}

One way to enhance the sensitivity is to maximize the contrast that is below unity because signal beams of different intensities interfere at the beam splitter~\cite{kolobov2017}.
However, in the Mandel configuration there is also the option to detect the intensity difference of both exits, namely $N_{-} = N_\text{s} - N_\text{s}^\prime$, which suffers no problem~\cite{miller2021} from interfering different intensities since $N_{-}= 2 n \sqrt{1 + n } \cos \phi$.
Here, the contrast can be viewed as perfect.
Without loss and following Appendix~\ref{app:Interference_patterns}, the variance can be written as $\operatorname{Var}(N_-) = N_+ + N_-^2 $, where also the sum $N_{+} =N_\text{s} + N_\text{s}^\prime$ of both exits contributes.
Consequently, we find the phase uncertainty obtained from such a measurement
\begin{align}
    \sigma^2_- = \frac{N_{+} + N_{-}^2}{(\partial N_{-}/\partial \phi)^2}.
\end{align}
Because the sum of the intensities $N_{+} = n (n+2)$ is independent of the interferometer phase and scales with $n^2$, the variance will never vanish.
Therefore, the only option to minimize the phase uncertainty is to work at mid fringe $\phi=\pi/2$, where $N_- =0$, while $\partial N_-/\partial \phi$ is maximal.
In fact, an optimization procedure finds exactly this phase and we obtain the phase uncertainty
\begin{align}
\label{s2}
    \left.\sigma^2_-\right|_{\pi/2} =\frac{1}{4n}\frac{n+2}{n+1}\overset{\text{hg}}{\longrightarrow} \frac{1}{4n},
\end{align}
which does not suffer from the severe sensitivity deterioration like the single exit detection scheme, but is still shot-noise limited.
This uncertainty of phase estimation coincides with the inverse of the quantum Fisher information associated to the signal modes, that is, with the idler photons undetected\footnote{
In fact, identifying $n=\sinh^2 r$ with gain $r$ in both crystals and using the relations $4n(n+1)=\sinh^2(2r)$ and $n(n+2)=\cosh^4 r-1$ gives rise to the inverse quantum Fisher information calculated in Ref.~\cite[Eq.~(30)]{kranias2025} without seed and external loss.
}.
Hence, differential detection at mid fringe constitutes an optimal measurement for the Mandel configuration.

For the lossless Yurke configuration it is optimal to work at destructive interference, whereas for the Mandel setup the working point lies at mid fringe.
In the former case, a two-mode squeezed state is generated and subsequently antisqueezed, while the latter displays conventional interference, which is characterized by the shot-noise limit.
This result is obtained by estimating the phase from the intensity difference~\cite{miller2021}, which is beneficial due to the unbalanced intensity of the two signal intensities, while still assuming equal gain.

Even though both configurations rely on squeezing and allow for bicolor imaging, the resulting sensitivities are quite different.
To gain further insights, we study a hybrid setup that allows for a continuous transition from the Yurke to the Mandel case.
For that, a part of the signal arm is coupled out of the Yurke interferometer and interferes on a beam splitter with the signal output of the second nonlinear medium.
This hybrid configuration has been effectively implemented in the low-gain regime~\cite{gemmell2024}.
Because it is necessary to achieve optimal sensitivity in the Mandel configuration, we resort to a detection of the differential intensity $N_{-}$ between both exits.

Following Fig.~\ref{fig:setups}\,(c), the reflectivities of the two beam splitters can be tuned by a parameter $\varrho$ in such a way that $\varrho=0$ reproduces the Yurke configuration, while $\varrho=1$ corresponds to the Mandel setup.
We choose the two phases of both setups to be the same and denote them by $\phi$.
The interference pattern as well as the variances and relevant transformations are given in Appendix~\ref{app:Interference_patterns}.

The resulting phase uncertainty is plotted as a function of the phase and of the tuning parameter $\varrho$ in Fig.~\ref{fig:hybrid} for a fixed photon number $n$.
Indeed, we observe a smooth transition from a single phase with minimal uncertainty in the Yurke case at the dark fringe $\phi=\pi$, towards optimal phases at two mid-fringe positions $\phi=\pi/2,3\pi/2$ in the Mandel case.
The optimal phases are denoted by white dots in the density plot.
In fact, for $\varrho\neq 0$ the phase uncertainty diverges at dark fringe $\phi =\pi$, which is a known feature of losses~\cite{marino2012,Xin2019},  where no perfect destructive interference is possible.

This feature demonstrates that the Mandel setup can be interpreted as the Yurke configuration including loss, which is recycled by a beam splitter before detection. 
The margin of the plot shows for each parameter setting of $\varrho$ the optimal phase and the associated phase uncertainty.
Here, we see that the phase uncertainty moves from the Heisenberg scaling to shot noise, but also demonstrates that gradual transitions between both regimes are possible.

\section{Phase sensitivity in a Yurke setup in the presence of loss}
\label{sec:loss_Yurke}
In the hybrid setup, we observed that recycling losses from the signal beam in the Yurke configuration, embedded within a Mandel scheme, causes a gradual transition from Heisenberg-limited to shot-noise scaling.
This observation underlines the role that optical losses play in determining the sensitivities of nonlinear interferometers.
Under realistic conditions, photon losses can occur in both arms of the interferometer, and the magnitude of loss may differ between the two paths.
Such a scenario is not expected to improve the sensitivity of the Mandel configuration.
However, it can reduce the capacity of the Yurke configuration to maintain Heisenberg-limited sensitivity, as has been demonstrated~\cite{marino2012} for the case of balanced losses in both arms.
In this Section we analyze the phase uncertainty of the Yurke configuration in the presence of loss.

Following Appendix~\ref{app:Interference_patterns} and assuming equal gain in both nonlinear media with vacuum input, the detected number of signal photons leaving the interferometer can be written as 
\begin{equation}
\label{Ns}
N_\text{s}=a+b\cos \phi,
\end{equation}
where
\begin{align}
\begin{split}
 a= &n \left(1+T_\text{s} \right)+n^2 \left( T_\text{s}+T_\text{i} \right)  \\
 b= & 2 n \left( n+1 \right) \sqrt{T_\text{s}  T_\text{i}} .
 \end{split}
\end{align}
Here, $T_\text{s}$ and $T_\text{i}$ are the overall losses in the signal/idler paths of the interferometer, $\phi$ is the phase to be estimated, and $n$ is the number of photons probing the sample located in the idler arm of the interferometer. 
The contrast of the interference pattern measured by varying the phase $\phi$ is $C=b/a$.

The uncertainty of phase estimation, using the propagation of errors equation, reads 
\begin{equation}
\label{sigma_Y_loss}
\sigma_\text{Y}^2=\frac{a(1+a)+b(1+2a) \cos \phi+b^2 \cos^2 \phi}{b^2 \sin^2 \phi}.
\end{equation}
The optimum phase $\phi_\text{min}$ that provides the minimum of the variance~\cite{Xin2019} given by Eq.~\eqref{sigma_Y_loss} can be calculated to be 
\begin{align}
\begin{split}
\label{eq:opt_uncertainty}
    \phi_\text{min} =\operatorname{cos}^{-1}&\left\{ - \frac{a(1 + a)+ b^2 }{b(1+2a)} \right. \\
    & \left. + \sqrt{ \left[ \frac{a(1+a) + b^2 }{b (1+2a)} \right]^2-1} \right\}.
\end{split}
\end{align}
Substituting the angle given in Eq.~\eqref{eq:opt_uncertainty} into the expression of the variance given by Eq.~\eqref{sigma_Y_loss}, we obtain that the minimum variance, for a given value of the parametric gain, is
\begin{align}
\begin{split}
\label{eq:sigma_Y_min}
    \left.\sigma_\text{Y}^2\right|_{\text{min}} =\, &\frac{a(1+a)-b^2}{2b^2} \\
    &+\frac{ \sqrt{\left( a^2 -b^2\right) \left[\left(1+a\right)^2 - b^2\right]} }{2 b^2}.
\end{split}
\end{align}

In order to determine if we can observe a quantum-enhanced sensitivity in phase estimation even under the presence of loss, it is helpful to make use of the concept of the normalized classical Fisher information ($F_\text{c}/n$).
The quantum nature of output signal photons in the Yurke setup corresponds to thermal light, so the classical Fisher information associated to the phase estimation protocol can be written as (see Appendix \ref{fisher_info_thermal})
\begin{equation}
F_\text{c}=1/\sigma_\text{Y}^2,
\end{equation}
where $\sigma_\text{Y}^2$ is given by Eq.~\eqref{sigma_Y_loss}. 
This connection implies that the mean photon number, i.\,e., the detected intensity, is an optimal phase estimator in the Yurke-type setup. 
Such a behaviour is expected since the probability distribution that describes signal photons (thermal light) belongs to the exponential family of probability distributions.

We consider as benchmark case for comparison the case of phase estimation using coherent light and the photon number $n$ probing the sample as a resource.
For a coherent state subject to the same loss $T_\text{i}$, or equivalently a Mach--Zehnder interferometer seeded by a coherent state, the Fisher information~\cite{kranias2025} $F_\text{c}^\text{coh}=4T_\text{i}n$ scales linearly with $n$.
A feature of phase estimation with a coherent light is thus that the normalized Fisher information $F_\text{c}^\text{coh}/n$ remains constant when increasing the number of photons $n$.
A quantum sensing protocol scheme for which $F_\text{c}/n$ increases with $n$ exhibits a local sub-shot-noise and by that a quantum-enhanced scaling.
On the contrary, if the protocol shows that $F_\text{c}/n$ decreases with $n$, it scales worse than shot noise. 
However, an actual sensitivity advantage requires a fair comparison of resources, that is, the same number of photons probing the sample and the same loss, and therefore additionally $F_\text{c}>F_\text{c}^\text{coh}$.

\begin{figure}[t!]
\centering
\includegraphics[width=\columnwidth]{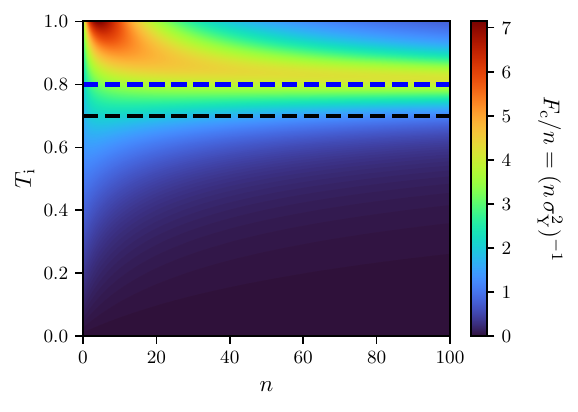}
\caption{
    Normalized classical Fisher information $F_\text{c}/n= 1/ (n\sigma_\text{Y}^2|_\text{min})$ at the optimal phase setting $\phi_\text{min}$ as a function of $n$ and $T_\text{i}$, assuming $T_\text{s}=0.8$ and equal gain.
    An increase of the normalized Fisher information with the number of probing photons implies a locally better than shot-noise scaling, while a constant behavior corresponds to shot-noise scaling.
    A decrease of the normalized Fisher information results in a deterioration of the phase sensitivity with the probing photon number, therefore scaling worse than shot noise. 
    We observe a resonance for  $T_\text{i}=T_\text{s}$.
    The cuts along the dashed lines at $T_\text{i}=0.7$ (black) and $T_\text{i}=0.8$ (blue) are displayed in Fig.~\ref{fig:YurkePhases}.
    }
    \label{fig:inverse}
\end{figure}

Using Eq.~\eqref{eq:sigma_Y_min}, we plot in Fig.~\ref{fig:inverse} the normalized classical Fisher information $F_\text{c}(\phi_\text{min})/n = 1/ (n \sigma^2_\text{Y}|_\text{min})$ at the optimal measurement point as a function of $n$ and $T_\text{i}$, assuming $T_\text{s}=0.8$.
In this plot, we observe a peak close to $T_\text{i}=T_\text{s}$, but even at this point the normalized classical Fisher information does not grow indefinitely, but saturates, indicating a shot-noise scaling in $n$.

\begin{figure}[t!]
\centering
\includegraphics[]{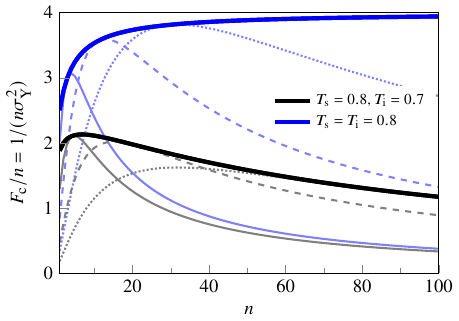}
\caption{
    Normalized classical Fisher information as a function of $n$, assuming equal gain.
    The blue lines correspond to a case with equal losses ($T_\text{s}=T_\text{i}=0.8$). 
    The thick blue solid line denotes the maximal classical Fisher information for each value of $n$, i.\,e. $F_\text{c}/n$ evaluated at $\phi_\text{min}$. 
    It approaches a constant value for high values of $n$, which corresponds to shot-noise scaling in the probing photon number. 
    The thin light blue lines are the normalized classical Fisher information for specific values of the phase, $\phi= 0.97 \pi, 0.95 \pi, 0.9 \pi$ (dotted, dashed, solid).
    The black lines corresponds to a case with non equal losses ($T_\text{s}=0.8$ and $T_\text{i}=0.7$).
    The thick black line denotes the maximal classical Fisher information, evaluated at $\phi_\text{min}$, whereas the thin gray lines corresponds to phases $\phi= 0.97 \pi, 0.95 \pi, 0.9 \pi$ (dotted, dashed, solid).
    The decrease of the normalized classical Fisher information signifies a scaling worse than shot noise in the probing photon number, whereas an increase corresponds to a local sub-shot-noise behavior.}
    \label{fig:YurkePhases}
\end{figure}

To study this behavior in more detail, we plot in Fig. \ref{fig:YurkePhases} the dependence of the normalized classical Fisher information for two cases that are representative of the general trend we have found exploring the whole parameter space.
Only for small photon numbers, we observe a local sub-shot-noise scaling in $n$ for both cases, identified from an increasing normalized classical Fisher information.
When losses inside the interferometer are equal ($T_\text{s}=T_\text{i} \equiv T$), the normalized classical Fisher information tends to a constant value for large values of $n$, as shown by the thick solid blue line, approaching shot-noise scaling in $n$.
When losses inside the interferometer are unbalanced ($T_\text{s} \neq T_\text{i}$), the normalized classical Fisher information decreases for large values of $n$ as shown by the thick solid black line.
This behavior indicates a scaling behavior in $n$ worse than shot noise.
In terms of the actual sensitivity and for the parameters chosen here, the maximal normalized Fisher information for equal loss exceeds the coherent reference $F_\text{c}^\text{coh}/n=4T_\text{i}$ for sufficiently large $n$, whereas for unequal loss it remains below this reference.

For a particular value of the phase (dotted, dashed, and solid thin lines) that does not coincide with $\phi_\text{min}$, we see in Fig. \ref{fig:YurkePhases} that there are regions of the parametric gain when the normalized classical Fisher information grows with increasing values of $n$, which corresponds to a locally improved scaling.
For large values of $n$, the variance decreases. 
Therefore, for a chosen phase $\phi$, eventually phase estimation for large enough values of $n$ scales worse than shot-noise in the probing photon number, since the normalized classical Fisher information decreases with increasing values of $n$. 
In this sense, only for the optimal phase $\phi_\text{min}$ one can approach shot noise at best.

In order to get further physical insight into the regime of high parametric gain, and thus large values of $n$, we expand Eq.~\eqref{eq:sigma_Y_min} for imperfect contrast\footnote{
Note that while the expansion is valid for generic but finite transmissions and contrast, one does not continuously recover Eq.~\eqref{eq:perfect_Yurke_Scaling} in the lossless limit $T_\text{s}=1=T_\text{i}$.
This is a consequence of the fact that the expansion in large $n$ and the limit of a lossless situation do not commute.
}
as a function of $1/n$ and we obtain
\begin{align}
\begin{split}
\label{eq:high-gain-Yurke}
    \left. \sigma^2_\text{Y}\right|_{\text{min}} \overset{\text{hg}}{\longrightarrow} & \frac{(T_\text{s}-T_\text{i})^2 }{4 T_\text{s}T_\text{i}} 
    + \frac{(T_\text{s}+T_\text{i}) (1-T_\text{i})}{2  T_\text{s}T_\text{i} n} \\
    &+ \frac{1 - T_\text{s} -3 T_\text{i} + 2 T_\text{s} T_\text{i} + 3 T_\text{i}^2}{4 T_\text{s} T_\text{i}n^2 } 
\end{split}
\end{align}
Equations~\eqref{sigma_Y_loss}, \eqref{eq:sigma_Y_min}, and \eqref{eq:high-gain-Yurke} are the main results of this section.
Equation~\eqref{sigma_Y_loss} is the general expression that determines the uncertainty of phase estimation for any phase $\phi$ and any parametric gain. 
Equation~\eqref{eq:sigma_Y_min} is the uncertainty of phase estimation when $\phi=\phi_\text{min}$, and thus gives the minimum variance for a given value of the parametric gain. 
Equation~\eqref{eq:high-gain-Yurke} is the minimum variance for the case of high parametric gain ($n \gg 1$) including loss, and it highlights the dependence of the variance of phase estimation on the value of the probing photon number $n$.

The first, constant contribution to Eq.~\eqref{eq:high-gain-Yurke} only arises if there is loss imbalance $T_\text{s}-T_\text{i} \neq 0$ and leads to a term that is not suppressed by increasing photon number.
Therefore, the value $(T_\text{s}-T_\text{i})^2/(4 T_\text{s}T_\text{i})$ poses the ultimate limit to the phase uncertainty, even if the phase is optimal.
However, this noise floor vanishes for balanced loss $T_\text{s}=T_\text{i}$, so that in this case the second contribution dominates.

\begin{figure}[t!]
\centering
\includegraphics[width=\columnwidth]{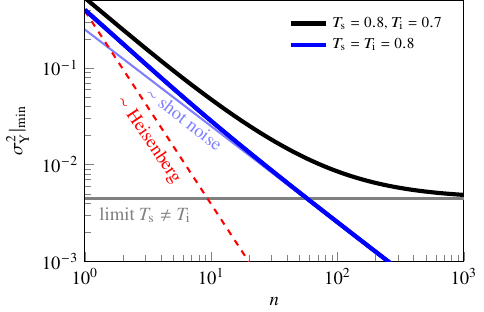}
\caption{
    Minimal phase uncertainty $\sigma_\text{Y}^2|_{\text{min}}$ as a function of $n$, assuming equal gain.
    For $T_\text{s}=0.8$ and $T_\text{i}=0.7$ (black), we observe a constant limit (gray line) in the high-gain regime, while for $T_\text{s}=T_\text{i}=0.8$ (blue) we observe shot-noise scaling in the probing photon number and with a loss-dependent prefactor.
    The dashed red line shows the $n^{-2}$ contribution, which becomes important for low photon numbers.
    The limits and scalings plotted in the figure correspond to the three terms of Eq.~\eqref{eq:high-gain-Yurke}.
    }
    \label{fig:scaling}
\end{figure}

The second contribution in Eq.~\eqref{eq:high-gain-Yurke} can be associated with shot noise due to its asymptotic scaling with $1/n$, but includes a loss-dependent prefactor and is the best one can achieve in a Yurke setup including loss.
However, for small photon numbers the third contribution becomes increasingly important.
Because the expansion ceases to be valid in this regime, this contribution only indicates a scaling better than shot noise, but does not imply Heisenberg scaling.
The optimal phase uncertainty from Eq.~\eqref{eq:sigma_Y_min} and these three contributions are plotted in Fig.~\ref{fig:scaling} for balanced loss and a loss imbalance.
Indeed, we observe that for unbalanced loss the phase uncertainty levels off to a constant noise floor, while for balanced loss we find asymptotic shot-noise scaling with a loss-dependent prefactor. 
However, while the exact curve for equal loss indeed scales better than shot noise, it does not approach Heisenberg scaling for small photon numbers.

We can get a clearer physical explanation of the different behavior of the minimum variance in the two regimes described above by rewriting Eq. (\ref{Ns}) as
\begin{align}
\begin{split}
 N_\text{s}= &n \left[ 1+T_s+2\sqrt{T_\text{s} T_\text{i}} \cos \phi \right]  \\
   &+n^2 \left[ T_\text{s}+T_\text{i}+2\sqrt{T_\text{s} T_\text{i}} \cos \phi \right]. \label{Ns2}
\end{split}
\end{align}
In the lossless case, the minimum variance is associated to the phase $\phi_\text{min}=\pi$ that yields $N_\text{s}=0$. 
Including loss, inspection of Eq.~\eqref{eq:opt_uncertainty} shows that in the high-gain regime, $\phi_\text{min} \rightarrow \pi$.
For this setting of $\phi$, the second term in Eq.~\eqref{Ns2}, scaling with $n^2$, vanishes if $T_\text{s} = T_\text{i}$.
However, this is not the case if $T_\text{s} \ne T_\text{i}$.
Therefore the variance of $N_\text{s}$, that is $N_\text{s} (N_\text{s}+1)$, is lower in the case with $T_\text{s} = T_\text{i}$ compared to the case with $T_\text{s} \ne T_\text{i}$. 
This reduced variance translates into better sensitivity in phase estimation for the case with $T_\text{s}=T_\text{i}$. 

In a realistic scenario, balanced loss is unfeasible.
We thus expect a sensitivity scaling worse than shot noise for such a Yurke setup.
The question arises whether the Yurke configuration is in fact, for \emph{realistic} experimental implementations, better suited than the Mandel configuration, which also allows for bicolor quantum imaging with shot-noise limited phase uncertainties.

\section{Comparison of Yurke and Mandel setups in the high-gain regime}\label{sec:YurkeVsMandel}

In a realistic experimental implementation of phase estimation with a Yurke configuration~\cite{topfer2022,kalashnikov2016,manceau2017a}, the likely scenario should correspond to the case $T_{s} \ne T_{i}$.
We have seen in Sec.~\ref{sec:loss_Yurke} that in the case of imbalanced losses inside the interferometer, going to the high-gain regime yields a sensitivity in phase estimation that scales even worse than shot noise in the probing photon number and has a loss-dependent prefactor.

In Sec.~\ref{sec:lossless}, we saw that in the lossless case, the Mandel configuration can provide a phase sensitivity scaling like shot noise when measuring the differential signal.
In this section, we will explore under which conditions, in the presence of loss, the Mandel configuration can still show such a behavior, thus surpassing the performance of the Yurke configuration in the high-gain regime.

We start the discussion by focusing on the detection of a single interferometer exit in the Mandel configuration, which even in the lossless case scales worse than shot noise.
The photon statistics is also thermal and the interference pattern has exactly the same form as in the Yurke case, but with a modified baseline and contrast, so that we can also resort to formally inserting the respective expressions for $a$ and $b$ into Eq.~\eqref{eq:opt_uncertainty} to obtain the optimal phase uncertainty.
Indeed, we find in Appendix~\ref{app:Interference_patterns} for equal gain the baseline $a= n (1 + T_\text{s} + n T_\text{i})/2$ and the parameter $b= n \sqrt{T_\text{s} T_\text{i}( 1+n)}$ that is connected to the contrast through $C=b/a$.
Because the contrast is smaller than unity, this experiment inhibits perfect destructive interference.
Indeed, we find in the high-gain limit
\begin{equation}
    \left.\sigma^2_\text{SM}\right|_{\phi_\text{min}} \overset{\text{hg}}{\longrightarrow} \frac{T_\text{i} n}{4 T_\text{s}} ,
\end{equation}
which is a generalization of the case without loss from Eq.~\eqref{s1}.
Not only is this scaling worse than shot noise, it is even increasing with increasing photon numbers $n$.
In contrast to the Yurke setup, the loss imbalance has no effect and this limit cannot be suppressed, as it will always be dominant for sufficiently high $n$.
Gain unbalancing may partially mitigate this issue.

Alternatively, measuring the photon-number difference $N_-$ between both exits of the interferometer~\cite{miller2021} provides another route that we study in the following.
Since our goal is to assess the performance of this specific intensity-based estimator, we employ moment propagation rather than the classical Fisher information of the full joint photon-number distribution.
With the variance of $N_-$ calculated in Appendix~\ref{app:Interference_patterns}, we find
the phase uncertainty
\begin{align} 
    \sigma^2_- = \frac{N_{+} + N_{-}^2 + 2(1-T_\text{i})T_\text{s}n^2}{(\partial N_{-}/\partial \phi)^2},
\end{align}
where the photon-number difference and sum of both exits take the explicit form
\begin{align}
\begin{split}
     N_- = &2 n \sqrt{(n+1) T_\text{i} T_\text{s}} \cos \phi \\
     N_+ = &n [1+ T_\text{s}+ n T_\text{i}].
\end{split}
\end{align}
This phase uncertainty is always minimized at mid-fringe for $\phi=\pi/2$, where $N_-=0$, so that no optimization depending on the transmittance is necessary.
In fact, for this phase setting $N_- =0$ and therefore the variance becomes minimal, since $N_+$ is independent of the phase.
At the same time, the slope of the photon-number difference in the denominator is maximal and we find
\begin{align}
    \left.\sigma^2_-\right|_{\pi/2} = \frac{1}{4T_\text{i}T_\text{s}n}\frac{1+T_\text{s}+n[T_\text{i}+2 T_\text{s} - 2T_\text{i}T_\text{s}]}{n+1},
\end{align}
which retains its shot-noise scaling in the high-gain regime, as the factor $T_\text{i}+2 T_\text{s} - 2T_\text{i}T_\text{s}$ never vanishes.
In the lossless limit $T_\text{s}=1=T_\text{i}$, this expression reduces to Eq.~\eqref{s2} and thus to the inverse of the quantum Fisher information of the signal modes when the idler remains undetected, so that internal loss only increases the phase uncertainty above this ideal bound.
Indeed, taking the high-gain limit leads to
\begin{align}
    \left.\sigma^2_-\right|_{\pi/2} \overset{\text{hg}}{\longrightarrow} \frac{1}{4n} \left(\frac{2}{T_\text{i}} + \frac{1}{T_\text{s}}-2\right),
\end{align}
which shows no exceptional point for equal loss $T_\text{i}=T_\text{s}$ and highlights that the asymptotic shot-noise scaling has a loss-dependent prefactor larger than unity.

\begin{figure}[t!]
\centering
\includegraphics[width=\columnwidth]{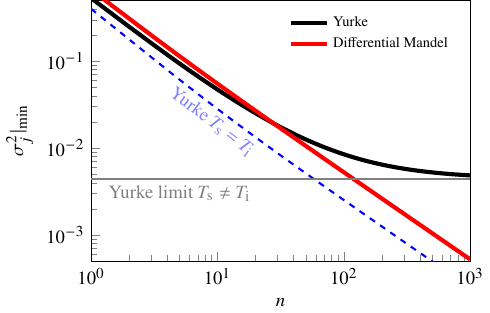}
\caption{
    Minimal phase uncertainty $\sigma_j^2|_{\text{min}}$ as a function of $n$, assuming equal gain.
    For the Yurke setup (Y, black) we observe for $\sigma_\text{Y}^2|_{\text{min}}$ a saturation (gray line) in the high-gain regime for $T_\text{s}=0.8$ and $T_\text{i}=0.7$ .
    In contrast, in the differential Mandel setup (M, red), there is no saturation and the sensitivity $\sigma_-^2|_{\pi/2}$ surpasses the one of the Yurke case, displaying clear shot-noise scaling with the probing photon number.
    For reference, the dashed blue line denotes the phase uncertainty of the Yurke setup for equal loss  $T_\text{s}=T_\text{i}=0.8$.
    \label{fig:comparison}
    }
\end{figure}

We compare this optimal phase uncertainty to that of the Yurke configuration in Fig.~\ref{fig:comparison}.
For unbalanced loss, the phase uncertainty of the Yurke setup saturates, whereas the Mandel setup always displays shot-noise scaling in the probing photon number and surpasses the Yurke sensitivity at sufficiently high photon numbers.
Even for balanced loss, the Yurke interferometer does not exhibit Heisenberg scaling for this range of parameters, providing at best only a fourfold improvement compared to the Mandel configuration because of the loss-dependent prefactors in the asymptotic scaling.
Since perfectly balanced loss is unlikely, the Mandel setup appears advantageous in high-gain quantum imaging.

\section{Conclusions}
Here, we have compared three imaging schemes for phase estimation based on nonlinear interferometers: 
a Yurke-type SU(1,1) interferometer, a Mandel-type induced coherence interferometer and a hybrid scheme that can interpolate between the two schemes.
In all cases we have considered unseeded configurations and equal gain in both nonlinear media. 
As measurement scheme, we have considered {\it easily accessible} intensity measurements and their first moment as estimators, avoiding more elaborate schemes such as coincidence measurements \cite{gerard2026} or homodyne detection \cite{anderson2017}. 
While such measurements can provide improved sensitivity for the Yurke configuration, they generally require more complex experimental implementations and are therefore beyond the scope of the present work, which focuses on simple and readily implementable intensity-based readout schemes.

From a practical perspective, our results provide design guidelines for quantum sensors based on nonlinear interferometers.
To achieve sub-shot-noise scaling, experiments should aim for moderate parametric gain, carefully balanced internal losses, and operation near destructive interference in a Yurke-type geometry.
In contrast, when large photon fluxes are needed or when asymmetric and sample-induced losses are unavoidable, a Mandel-type interferometer with differential detection offers a more robust, though shot-noise-limited, solution that can outperform the Yurke setup in the relevant parameter regime.

This trade-off between ultimate scaling and robustness directly affects potential applications of nonlinear interferometers in quantum-enhanced metrology and imaging, especially when the sample itself is a major source of loss.
The parameter-space studies and scaling laws developed here serve as a guide for choosing architectures and operating points in future experiments, and they encourage further research on engineered-loss platforms, alternative detection strategies, and multimode extensions aimed at recovering quantum-enhanced sensitivity under more realistic experimental conditions.

The noise floor of the Yurke configuration is determined by the loss imbalance $T_\text{s}-T_\text{i}$.
In principle, this limitation can be mitigated by deliberately introducing additional loss in the less lossy arm, thereby balancing the effective transmission coefficients.
However, such an approach requires precise control of the added attenuation, as fluctuations in absorption, spatial variations across pixels, and uncertainties in the transmission coefficients may ultimately limit the achievable sensitivity.

Looking ahead, our analysis can be extended in several directions that are relevant for practical quantum sensors.
Beyond the internal losses considered here, external detection loss and detector noise can be incorporated using quantum-imaging distillation and detection-loss-tolerant schemes~\cite{schaffrath2024a,fuenzalida2023}, yielding a full error budget from the nonlinear crystals to the camera.
Another promising direction is to investigate gain unbalancing as a strategy for mitigating the effects of loss~\cite{giese2017,manceau2017,kolobov2017}.
While the present analysis in the main body of our work assumes equal gain in both nonlinear media, gain-unbalanced configurations are in principle included in the general expressions of Appendix~\ref{app:Interference_patterns} and provide an interesting basis for future research.

Our conclusions are restricted to unseeded configurations.
Coherent seeding can enhance the phase sensitivity of nonlinear interferometers and may modify the relative performance of different detection schemes~\cite{kranias2025,miller2021}.
However, including seeding introduces additional optimization parameters, such as seed amplitudes and phases, and requires a separate analysis beyond the scope of the present comparison.

It will also be important to revisit these architectures from the perspective of quantum Fisher information~\cite{kranias2025}.
Yurke and Mandel configurations have the same ultimate quantum Fisher information if both signal and idler modes are accessible, but realistic measurement schemes access very different fractions of it once the idler remains undetected or loss and noise are present.
Quantifying how alternative readout strategies, such as balanced detection, homodyne or truncated SU(1,1) protocols, approach these bounds in realistic multimode imaging scenarios will guide the choice of architectures for bicolor quantum imaging, nonlinear interferometric spectroscopy, and tomography.

For the Mandel configuration without loss, differential detection constitutes an optimal measurement, since its phase uncertainty saturates the quantum Fisher information of the signal modes when the idler remains undetected.
We expect this optimality to persist in the presence of internal loss, since the structure of the reduced state of the signal modes remains unchanged.
However, establishing this property rigorously requires either the quantum Fisher information of the lossy signal modes or, restricted to photon-number measurements, the classical Fisher information of the full joint photon-number distribution at both output ports, both of which go beyond the moment-based analysis of this work.

In summary, our results provide guidance on what can be achieved with {\em simple} and {\em experimentally easily accessible} intensity measurements in the Yurke-type and Mandel-type setups, the two configurations most widely used in nonlinear interferometry.
We identify the configuration that provides the best sensitivity for phase estimation under different gain and loss regimes, and the conditions under which sub-shot-noise scaling can be achieved with intensity measurements.

\section*{Acknowledgments}
The authors thank P. Schach and D. Derr for stimulating discussions.
D.F.U., A.A.P. and J.P.T. acknowledge support from the project NOVISLIGHT (PID2023-149780NB-I00) funded by Ministerio de Ciencia, Innovaci\'on y Universidades (Proyectos de generaci\'on de conocimiento 2023).
This work was partially funded by CEX2024-001490-S [MICIU/AEI/10.13039/501100011033]. G.J.M., R.B.P. and I.A.W. acknowledge support from the National Research Council Canada (project QSP 062-2) and the UK Research and Innovation Future Leaders Fellowship (project: MR/W011794/1), Department of Science, Innovation \& Technology Tactical Fund.
Support is acknowledged from the German Federal Ministry of Education and Research (BMBF) within the funding program ``quantum technologies -- from basic research to market'' with contract number 13N16496 (QUANCER). 
Moreover, this work is also funded by the Deutsche Forschungsgemeinschaft (DFG, German Research Foundation) -- Projektnummer 552245798.

\section*{Data availability}

All data and code required to reproduce the figures presented in this paper are publicly available in an online repository at \url{https://doi.org/10.5281/zenodo.21703011}, based on \url{https://github.com/Cristofero1/below-shot-noise-phase-estimation}.

\section*{Author contributions}
\textit{Conceptualization}: IAW, MG, JPT, EG; 
\textit{Formal analysis}: CO, GJM, FF;
\textit{Funding acquisition}: IAW, MG, JPT;
\textit{Investigation}: CO, GJM, FF, JPT;
\textit{Methodology}: CO, GJM, FF, DFU, AAP, RBP, IAW, MG, JPT, EG;
\textit{Supervision}: IAW, MG, JPT, EG;
\textit{Validation}: DFU, AAP , RBP, IAW, MG, JPT, EG;
\textit{Visualization}: CO, JPT, EG;
\textit{Writing -- original draft}: CO, GJM, JPT, EG;
\textit{Writing -- review \& editing}: CO, GJM, FF, DFU, AAP, RBP, IAW, MG, JPT, EG. \newline
A large language model was used for language editing and grammar correction.

\appendix
\section{Interference patterns}
\label{app:Interference_patterns}

\subsection{Interferometer output operator}
In this appendix, we derive the bosonic operators that account for the evolution of a photon through the setup.
To describe parametric down-conversion, we use Bogoliubov transformations that have coupling matrices that are elements of the SU(1,1) group.
For our two nonlinear media $j=A,B$, they take the form
\begin{equation}
    \hat{b}_\text{s/i}= u_j \hat{a}_\text{s/i} + v_j \hat{a}^\dagger_\text{i/s},
\end{equation}
where $\hat{b}_\text{s/i}$ is the output annihilation operator of that particular medium in signal/idler mode and $\hat{a}_\text{s/i}$ is the respective input annihilation operator, which of course fulfill the usual bosonic commutation relations.
The complex coefficients $u_j$ and $v_j$ increase exponentially with parametric gain, but also include the phase of the pump laser at interaction, as parametric amplification is a phase-sensitive process.
They fulfill the property $|u_j|^2-|v_j|^2 = U_j - V_j=1$, which corresponds to the hyperbolic identity, so that the coefficients can be parametrized by hyperbolic functions that scale exponentially with gain.

The loss on each arm $j=\text{s},\text{i} $ is included by conventional SU(2) beam-splitter matrices, namely
\begin{equation}
    \hat{b}_j = t_j \hat{a}_j+ r_j \hat{l}_j ,
\end{equation}
where the bosonic operators $\hat{l}_j$ describe vacuum input modes of the respective loss channels.
The complex transmission coefficients fulfill the relation $|t_j|^2 + |r_j|^2 = T_j + R_j =1$ and imply a transmittance $T_j$ on arm $j$.
This transmittance may include the transmittance of the object used in an imaging setup, but also any other loss on that arm.
Moreover, the phase acquired on each arm can be included as the phase of $t_j$, so that the phase shift induced by the imaged object is accounted for.

To find the transformations for the different setups, one can step by step follow the sketches in Fig.~\ref{fig:setups} and insert the corresponding SU(1,1) and SU(2) transformations.
The Yurke configuration is characterized by the output operator 
\begin{align}
\label{eq:operator_yurke}
    \hat{b}_\text{Y} = \alpha_\text{Y} \hat{a}_{\text{s}A} +  \kappa_\text{Y} \hat{l}_\text{s} + \beta_\text{Y} \hat{a}_\text{i}^\dagger +\lambda_\text{Y} \hat{l}_\text{i}^\dagger,
\end{align}
in the signal mode, which effectively describes the transformation of the entire interferometer. 
Here, the annihilation operators $\hat{a}_{\text{s}A}$ and $\hat{a}_\text{i}$ denote vacuum input at the first nonlinear medium $A$, while the annihilation operators operators $\hat{l}_\text{s}$ and $\hat{l}_\text{i}$ are the vacuum inputs associated with the loss channels in the signal and idler arm, respectively.
The complex path coefficients contain all the information of the interferometer, that is, the gain, loss, sample properties, and phases, and are given by
\begin{subequations}
    \begin{align}
    \alpha_\text{Y} &= u_B t_\text{s} u_A + v_B t_\text{i}^\ast v_A^\ast  \text{ and } \kappa_\text{Y} = u_B r_\text{s} \\
    \beta_\text{Y}& = u_B t_\text{s} v_A + v_Bt_\text{i}^\ast u_A^\ast \text{ and }
    \lambda_\text{Y} = v_B r_\text{i}^\ast .
\end{align}
\end{subequations}

For the Mandel configuration, the signal mode of medium $A$ is not used as an input for medium $B$.
Hence, an additional (vacuum) input channel associated with operator $\hat{a}_{\text{s}B}$ has to be included into the description. 
Moreover, the two signal modes interfere on a 50:50 beam splitter through the SU(2) transformation 
$\hat{b}_\text{M}^{(\pm)} = (\hat{b}_{\text{s}A} \pm \hat{b}_{\text{s}B})/\sqrt{2} $.
As a consequence, we find two output operators after the beam splitter, namely
\begin{align}
\begin{split}
\label{eq:operator_mandel}
    \hat{b}^{(\pm)}_\text{M} =
    &\,
    \alpha^{(\pm)}_\text{M} \hat{a}_{\text{s}A} 
    + \gamma^{(\pm)}_\text{M} \hat{a}_{\text{s}B} 
    + \kappa_\text{M} \hat{l}_\text{s} \\
    &
    + \beta^{(\pm)}_\text{M} \hat{a}_\text{i}^\dagger 
    +\lambda^{(\pm)}_\text{M} \hat{l}_\text{i}^\dagger, 
\end{split}
\end{align}
with path coefficients
\begin{subequations}
    \begin{align}
    \alpha_\text{M}^{(\pm)} &= \frac{t_\text{s} u_A \pm v_B t_\text{i}^\ast v_A^\ast}{\sqrt{2}} \text{ , }     \gamma_\text{M}^{(\pm)}  = \frac{\pm u_B }{\sqrt{2}} \text{ , }     \kappa_\text{M} =  \frac{r_\text{s}}{\sqrt{2}}\\
    \beta_\text{M}^{(\pm)} &= \frac{ t_\text{s} v_A \pm v_B t_\text{i}^\ast u_A^\ast}{\sqrt{2}} \text{ and }
    \lambda_\text{M}^{(\pm)} = \pm\frac{v_B r_\text{i}^\ast}{\sqrt{2}}.
\end{align}
\end{subequations}

Modifying the Mandel setup such that the lost fraction of intensity of the signal mode couples back into the second nonlinear medium leads to a hybrid between the Mandel and Yurke configuration, where the path coefficients are altered by the modified seed of the second medium.
We therefore arrive at
\begin{align}
\label{eq:operator_hybrid}
    \hat{d}^{(\pm)}_\text{H} = \alpha^{(\pm)}_\text{H} \hat{a}_{\text{s}A}+ \gamma^{(\pm)}_\text{H} \hat{a}_{\text{s}B} + \beta^{(\pm)}_\text{H} \hat{a}_\text{i}^\dagger,  
\end{align}
for a configuration without loss, where we have defined the path coefficients
\begin{subequations}
    \begin{align}
    \alpha_\text{H}^{(+)} = -r r_\text{s}^\ast u_A +t \alpha_\text{Y}&\text{ , }
    \alpha_\text{H}^{(-)} = -t^\ast r_\text{s}^\ast  u_A -r^\ast \alpha_\text{Y}
    \\
    \beta_\text{H}^{(+)} = -r r_\text{s}^\ast  v_A +t \beta_\text{Y}
    &\text{ , }
    \beta_\text{H}^{(-)} = -t^\ast r_\text{s}^\ast  v_A -r^\ast \beta_\text{Y} 
    \\
    \gamma_\text{H}^{(+)}  = t  u_B r_\text{s} + r t_\text{s}^\ast 
    &\text{ , } 
    \gamma_\text{H}^{(-)}  = -r^\ast  u_B r_\text{s} + t^\ast t_\text{s}^\ast.
\end{align}
\end{subequations}
We now introduce the tuning parameter $\varrho = 1- T_\text{s}$ to tune the fraction that couples out of the Yurke setup.
At the same time, we choose the transmittance of the final beam splitter to be $|t|^2 = 1- \varrho /2$, in agreement with Fig.~\ref{fig:setups}.
These choices lead for $\varrho = 0$ to the Yurke and for $\varrho = 1$ to the Mandel configuration.

\subsection{Interference signal}

In contrast to configurations that use a coherent input to boost the performance of the interferometers~\cite{plick2010}, we assume no seed, that is, vacuum input.
In this case, the detected photon numbers are always of the form $|\beta_j|^2+|\lambda_j|^2$ for all three considered setups, where $|\lambda_j|^2$ can be interpreted as vacuum noise introduced by loss.

As reported before~\cite{giese2017}, we observe for the Yurke configuration that the detected signal photons can be described by the interference pattern $ N_\text{s} = a + b \cos \phi$, with
\begin{subequations}
\begin{align}
    a &= T_\text{s} V_A + V_B + (T_\text{s}+T_\text{i}) V_A V_B\\
    b&= 2 \sqrt{T_\text{s} T_\text{i} V_A V_B (1+V_A)(1+V_B)} .
\end{align}
\end{subequations}
Here, $a$ corresponds to the baseline of the pattern, the contrast is determined by $C=b/a$, and the interferometer phase has been defined as $\phi =\operatorname{arg} (t_\text{s}t_\text{i} v_A v_B^\ast u_A u_B)$.
For equal gain, $V_A=V_B=n$ corresponds to the number of photons generated in nonlinear medium $A$ and is used to determine whether we observe sub-shot-noise sensitivities.

For the Mandel configuration, the two exits show the photon numbers
\begin{subequations}
\begin{align}
    N_\text{s}&= \big\langle\hat{b}^{(+)\dagger}_\text{M} \hat{b}^{(+)}_\text{M}\big\rangle = a + b \cos \phi \\
    N_\text{s}'&= \big\langle\hat{b}^{(-)\dagger}_\text{M} \hat{b}^{(-)}_\text{M}\big\rangle = a - b \cos \phi,
\end{align}    
\end{subequations}
with the interferometer phase $\phi =\operatorname{arg}( t_\text{s}t_\text{i} v_A v_B^\ast u_A) $, which effectively corresponds to the phase observed for the Yurke setup as $u_B$ can be assumed as real.
Here, we find the parameters
\begin{subequations}
\begin{align}
    a &= (T_\text{s}V_A+V_B+T_\text{i}V_AV_B)/2 \\
    b&= \sqrt{T_\text{s} T_\text{i} V_A V_B (1+V_A)},
\end{align}
\end{subequations}
as a slight generalization of established results~\cite{kolobov2017}.

Next, we turn to the hybrid configuration without loss and rewrite the tuning parameter as $\varrho = 1-\tau$. 
Because the setup effectively corresponds to three-path interference~\cite{gemmell2024}, in general a beating occurs.
However, we can identify the phases associated with the Mandel and Yurke setups, as above, and choose in the following $\phi_\text{M}=\phi_\text{Y}=\phi$.
Here, the phase of the Mandel setup includes an additional shift of $\pi$ imprinted upon reflection by the beam splitter that couples light out of the Yurke setup.
This phase shift is necessary so that the Mandel result can be recovered without the two exits changing their role.
With the identification $T_\text{s} = \tau$ and $|t|^2 = (1+\tau)/2$, we find for the sum of the photon numbers of both exits
\begin{equation}
         N_+ =n\left[n+2 + n\tau+ 2(1+n)\sqrt{\tau}\cos\phi\right],
\end{equation}
and a photon-number difference in the usual form $N_- = a + b\cos \phi$, however, with
\begin{subequations}
\begin{align}
        a&=  2n \varrho \sqrt{(n+1)(\tau+\tau^2)}+n^2 \tau + n(2+n)\tau^2\\
        b&=2n\bigg[ (1+n)\tau^{3/2}+\varrho\sqrt{(1+n)(1+\tau)}\bigg].
 \end{align}
\end{subequations}

\subsection{Variances of detected photon numbers}

With the help of basic bosonic commutation relations, we bring the variance to normal order, that is, $\operatorname{Var}(N_\text{s}) = \big\langle \hat{b}_j^\dagger \hat{b}_j \hat{b}_j^\dagger \hat{b}_j \big\rangle - N_\text{s}^2 =\big\langle \hat{b}_j^{\dagger 2} \hat{b}_j^2  \big\rangle + N_\text{s}- N_\text{s}^2 $.
The second-order correlation function, when acting on the vacuum input, can then be expressed by 
\begin{equation}
    \big\langle \hat{b}_j^{\dagger 2} \hat{b}_j^2  \big\rangle = 2 (|\beta_j|^4 + |\lambda_j|^4) + 4 |\beta_j|^2 |\lambda_j|^2= 2 N_\text{s}^2.
\end{equation}
As a consequence, we find the variance
\begin{equation}
    \operatorname{Var}(N_\text{s}) = N_\text{s} [1+N_\text{s}],
\end{equation}
in agreement with thermal statistics.
The same calculation holds for the other exit and the variance of $N_\text{s}^\prime$.

To calculate the variance of the photon-number difference $N_-=N_\text{s} -N_\text{s}^\prime$ in the Mandel setup, we first identify
$\operatorname{Var}(N_-) = \operatorname{Var}(N_\text{s}) +\operatorname{Var}(N_\text{s}^\prime) - 2 \operatorname{Cov}(N_\text{s} ,N_\text{s}^\prime)$, where we introduce the covariance
\begin{equation}
\operatorname{Cov}(N_\text{s} ,N_\text{s}^\prime) = \frac{\big\langle \hat{b}_\text{M}^{(+)\dagger }\hat{b}_\text{M}^{(-)\dagger } \hat{b}_\text{M}^{(+) }\hat{b}_\text{M}^{(-) }  \big\rangle + \text{c.c.}}{2}- N_\text{s} N_\text{s}^\prime,
\end{equation}
already normally ordered.
When the respective operators act on the input vacuum state, we find for the cross-correlation function the expression
\begin{equation}
    \big\langle \hat{b}_\text{M}^{(+)\dagger }\hat{b}_\text{M}^{(-)\dagger } \hat{b}_\text{M}^{(+) }\hat{b}_\text{M}^{(-) }  \big\rangle = 2 N_\text{s}N_\text{s}^\prime - V_B V_A T_\text{s} (1-T_\text{i}).
\end{equation}
With that, the variance of the photon-number difference reduces to
\begin{equation}
    \operatorname{Var}(N_-) = N_+ + N_-^2 + 2 V_B V_A T_\text{s} (1-T_\text{i}),
\end{equation}
with $N_+ = N_\text{s} + N_\text{s}^\prime $.
For the lossless hybrid setup, we also find the respective limit $\operatorname{Var}(N_-) = N_+ + N_-^2$.

\section{Classical Fisher information associates to thermal light}
\label{fisher_info_thermal}
The photon-number probability distribution associated to thermal light is
\begin{equation}
\label{distribution_pm}
p_m=\frac{1}{1+N_\text{s}(\phi)}\, \left[ \frac{N_\text{s}(\phi)}{1+N_\text{s}(\phi)}\right]^m,
\end{equation}
where $m$ is the number of photons and $\phi$ is the phase to be estimated. $N_\text{s}(\phi)$ is the mean value of the probability distribution, and the variance is $N_\text{s}(\phi) \left[ 1+N_\text{s}(\phi) \right]$.

The classical Fisher information associated to a probability distribution $p_m$ can be written as
\begin{equation}
\label{fisher10}
F_\text{c}(\phi)
=-\sum_m p_m\, \left( \frac{\partial^2 \ln p_m}{\partial \phi^2} \right).
\end{equation}
From Eq.~\eqref{distribution_pm}, we obtain that
\begin{align}
\begin{split}
 \frac{\partial^2 \ln p_m}{\partial \phi^2}= \left\{ \frac{1+m}{\left[ 1+N_\text{s} \right]^2}-\frac{m}{N_\text{s}^2 } \right\} \left[ \frac{\partial N_\text{s}}{\partial \phi} \right]^2 \\
 \label{fisher20} + \left\{ - \frac{1+m}{ 1+N_\text{s}}+\frac{m}{N_\text{s} } \right\} \frac{\partial^2 N_\text{s}}{\partial \phi^2}.
 \end{split}
\end{align}
Making use of $\sum_m p_m=1$ and $\sum_m m\, p_m=N_\text{s}(\phi)$, substitution of Eq.~\eqref{fisher20} into Eq.~\eqref{fisher10} yields
\begin{equation}
F(\phi)=\frac{1}{N_\text{s}(\phi) \left[ 1+N_\text{s}(\phi) \right]}  \left[ \frac{\partial N_\text{s}(\phi)}{\partial \phi} \right]^2.
\end{equation}
 We have thus demonstrated that the classical Fisher information is $F(\phi)=1/\sigma_{Y}^2(\phi)$, where $\sigma_{Y}^2(\phi)$ is obtained using the propagation of errors equation.
\vspace{1em}

\bibliographystyle{quantum}
\bibliography{references}

\end{document}